\documentclass[twocolumn,showpacs,aip,apl,preprintnumbers,amsmath,amssymb,superscriptaddress,reprint]{revtex4-1}

\usepackage{graphicx}
\usepackage{dcolumn}
\usepackage{bm}

\usepackage{color}

\begin{document}


\title{High-Transconductance Graphene Solution-Gated Field Effect Transistors}

\author{L. H. Hess}
\author{M. V. Hauf}
\author{M. Seifert}
\affiliation{Walter Schottky Institut, Technische Universit\"{a}t M\"{u}nchen, Am Coulombwall 4, 85748 Garching, Germany}
\author{F. Speck}
 
\author{T. Seyller}
\affiliation{Friedrich-Alexander-Universit\"at Erlangen-N\"urnberg, Erwin-Rommel-Stra\ss e 1, 91058 Erlangen, Germany }
\author{M. Stutzmann}
\author{I. D. Sharp}
\author{J. A. Garrido}
\email{garrido@wsi.tum.de} 
\affiliation{Walter Schottky Institut, Technische Universit\"{a}t
M\"{u}nchen, Am Coulombwall 4, 85748 Garching, Germany}

\date{\today}

\begin{abstract}
In this work, we report on the electronic properties of solution-gated field effect transistors (SGFETs) fabricated using large-area graphene. Devices prepared both with epitaxially grown graphene on SiC as well as with chemical vapor deposition grown graphene on Cu exhibit high transconductances, which are a consequence of the high mobility of charge carriers in graphene and the large capacitance at the graphene/water interface. The performance of graphene SGFETs, in terms of gate sensitivity, is compared to other SGFET technologies and found to be clearly superior, confirming the potential of graphene SGFETs for sensing applications in electrolytic environments.
\end{abstract}


\maketitle
\begin{figure}
	\centering
		\includegraphics{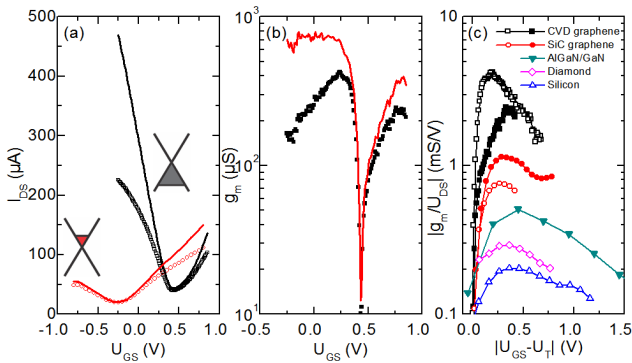}
		\caption{\label{fig:TRT} (a) Drain-source current vs. gate voltage for SiC (red circles) and CVD (black squares)  graphene SGFETs (Channel length 20 $\mu$m). The solid lines correspond to the expected currents with the contribution of the access resistance taken into account. The small insets indicate the positions of the Fermi levels for ungated SiC (left) and CVD (right) graphene. (b) Transconductance of a CVD graphene SGFET. The symbols represent the experimental results, whereas the line shows the internal transconductance calculated by considering the access resistance. (c) Normalized  transconductance of SGFETs based on graphene  and  other  materials, shifted by the transistor threshold  voltage $U_T$  ($U_D$ for graphene  devices).  Open symbols represent p-type and closed symbols n-type devices. All transistors have the same channel length-to-width ratio.
		}
\end{figure}

Due to its extraordinary chemical and  electrochemical  properties,\cite{Tang09} graphene is a promising  candidate  for sensing in electrolyte environments. 
To date, most reports in this area concern the use of graphene and graphene-related materials as electrodes for sensing applications.\cite{Lu09}  However, the implementation of transistor-based sensor concepts offers several advantages, such as intrinsic signal amplification and facile integration with microelectronic  circuits. For the particular case of graphene solution-gated  field effect transistors (SGFETs), the high mobilities reported for electrons and holes suggest devices with large transconductances and, thus, high sensitivities.
In addition to the detection of electrolyte properties, such as pH or ionic strength,\cite{Ang08} graphene SGFETs are suitable for the investigation of  more complex systems and phenomena, including the electrical activity of living cells.\cite{Cohen-Karni10} Recently, graphene  SGFETs  have been realized using exfoliated  graphene flakes\cite{Heller10} and epitaxially  grown graphene on SiC.\cite{Ang08,Dankerl10,Ristein10} For biological applications, arrays of $\mu$m-sized transistors are advantageous,  e.g.  for the investigation  of cellular communication in neural networks. In this respect,  large-scale graphene sheets grown by thermal decomposition of SiC\cite{Berger06} or by chemical vapor deposition (CVD)\cite{Li09} are of special interest. In this work, we demonstrate the fabrication of arrays of SGFETs using epitaxial graphene as well as CVD graphene. In comparison to devices based on competing material systems, such as Si, diamond, or AlGaN/GaN, graphene SGFETs exhibit superior transconductances, which as we show arises from the combined contribution of the high carrier mobilities in graphene and the large capacitance of the graphene/electrolyte interface.\\
\begin{figure}
	\centering
		\includegraphics{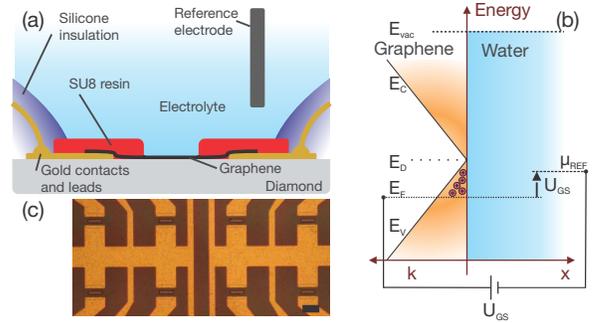}
		\caption{\label{fig:SCH}(a) Layout of a graphene SGFET showing the Ag/AgCl reference electrode used to control the graphene/electrolyte interfacial potential. (b) Modulation of the charge carrier density by the electrolyte gate: the applied gate voltage shifts the Fermi level in graphene below (shown) or above the Dirac point ($E_D$), defining the density and type of charge carriers. (c) Optical micrograph of an SGFET array with large access regions (scale bar corresponds to 50 $\mu$m).
}
\end{figure}
CVD graphene was grown on copper foil at $1000\,^{\circ}{\rm C}$ in a CH$_4$/H$_2$/Ar atmosphere, as reported previously.\cite{Li09} Following growth, standard methods for selective Cu etching and material transfer\cite{Mattevi11} were used to place graphene 
onto insulating oxygen-terminated single-crystalline  diamond, which was pre-patterned with Ti/Au metal  contacts. The active areas of the graphene transistors, with gate widths of 40 $\mu$m and lengths of either 16 $\mu$m or 26 $\mu$m, were defined by UV-photolithography and oxygen plasma etching. Finally, the samples were annealed in vacuum at $500\,^{\circ}{\rm C}$ to remove residual contamination. SGFETs were also prepared from epitaxial graphene on the Si-face of SiC by thermal decomposition at $1600^{\circ}{\rm C}$ under argon atmosphere.\cite{Emtsev09} For these devices, the electrical leads were fabricated by local plasma oxidation of the graphene  and evaporation of Ti/Au. For both  CVD and SiC graphene devices, the metal contacts  and leads were insulated  from the electrolyte by a chemically stable photoresist (SU8). The partial overlap of the SU8 layer with the graphene in the transistor active area (3 $\mu$m on both the drain and source region) resulted in devices with effective gate lengths of 10 $\mu$m and 20 $\mu$m. The experiments were performed in a 10 mM sodium phosphate-buffered electrolyte adjusted  to a total  ionic strength  of 50 mM by adding NaCl. The transistors  were biased in a two-electrode setup using a Ag/AgCl-wire  as the reference electrode (see Fig. \ref{fig:SCH}(a)).  Drain-source  and  gate-source biases were applied using two source meters.\\
Fig. \ref{fig:TRT}(a) shows the experimental  results  of the drain-source  current $I_{DS}$ as function of the gate voltage $U_{GS}$ at a drain-source  voltage $U_{DS}$ of 100 mV. The current modulation shown in Fig. \ref{fig:TRT}(a) results from the gate-induced shift of the Fermi level (Fig. \ref{fig:SCH}(b)) which controls the density and type of charge carriers. The current minimum, which corresponds to the Dirac point $U_D$, is observed at $U_{GS}$ = +430 mV for CVD graphene and $U_{GS}$ = -250 mV for SiC graphene. The  different gate potentials of the Dirac point can be  explained  by substrate-induced doping. The Fermi level of ungated graphene can be estimated considering the intrinsic work function of undoped graphene ($4.6\pm0.1$ eV\cite{Yu09}), the potential of the Ag/AgCl reference of 4.7 eV below the vacuum level, and the Dirac point. 
Fig. \ref{fig:TRT}(a) shows that the SiC graphene is n-type doped in good agreement with previous results,\cite{Emtsev09} with $E_F-E_D\approx100$ \nolinebreak meV. On the other hand, a p-type doping, with $E_F-E_D\approx-500$ \nolinebreak meV, is obtained for the CVD-grown graphene. A similar $E_F$ position has been observed for SiO$_2$ and is attributed to the doping effect of a water layer underneath the graphene.\cite{Wehling08}\\
\\
For  both the electron ($U_{GS}>U_D$) and  the hole regime ($U_{GS}<U_D$), a deviation from a $I_{DS}-U_{GS}$ linear dependence can  be observed for $|U_{GS}-U_D| >300$ mV. However, the observed behavior is not symmetric for electrons and holes. For SGFET devices on CVD graphene, the slope of the $I_{DS}-U_{GS}$ curve is clearly higher for $U_{GS}>U_D$, whereas on SiC graphene devices the opposite effect is observed. Fig. \ref{fig:TRT}(b) shows the transconductance $g_m$ of the CVD graphene SGFET in Fig. \ref{fig:TRT}(a), calculated as the derivative of $I_{DS}$ with respect to $U_{GS}$. This parameter is of special interest for sensor applications as it specifies the response of the transistor, i.e. the current response to a small modulation of the gate voltage, and thus its sensitivity. 
Maxima of $|g_m|$ are observed in both the electron and hole regimes, reaching values of 250 $\mu$S (electrons) and 420 $\mu$S (holes) for CVD graphene, and 110 $\mu$S (electrons) and 80 $\mu$S (holes) for SiC graphene.\\
The flattening of the current response to gate voltage observed away from the Dirac point is a consequence of the access regions. As shown in Fig. \ref{fig:SCH}(a), the graphene conductivity in the region covered with SU8 cannot be controlled by the electrolytic gate. These regions will be referred to as access regions, and can be modeled by a gate independent access resistance $R_{AR}$, which is extracted by comparing transistors with different effective channel lengths but identical access regions. By removing $R_{AR}$ from the total resistance of the device, the "internal" conductivity of the transistor can be calculated (solid lines in Fig. \ref{fig:TRT}(a)). The reduced current response is still present in the CVD G-SGFET for $U_{GS}>U_D$, and in the SiC G-SGFET for $U_{GS}<U_D$, suggesting a difference in electron and hole transport for both types of devices. This asymmetry can be explained by considering the contribution of the fixed Fermi level in the graphene below the SU8. Depending on the carrier type in the open channel, a p-n junction can form at the boundary with the access region, resulting in an additional resistance due to restricted carrier injection. 
Based on the discussion of the substrate-induced doping, the ungated regions of the graphene transistors under the SU8 layer are expected to be p-type and n-type doped for the CVD and SiC graphene, respectively. Therefore, in the case of CVD graphene, a restriction of the electron conduction is expected when the graphene channel is biased in the n-type regime ($U_{GS}>U_D$).
For SiC graphene, on the other hand, the access regions are expected to be n-type, leading to the observed restricted hole conduction for $U_{GS}<U_D$. A similar asymmetry in the conductivity was previously observed and attributed to the doping effect of metal contacts on the underlying graphene.\cite{Huard08} In our case, however, the local doping effect is expected to be mostly caused by the access regions.\\
\begin{figure}
	\centering
		\includegraphics{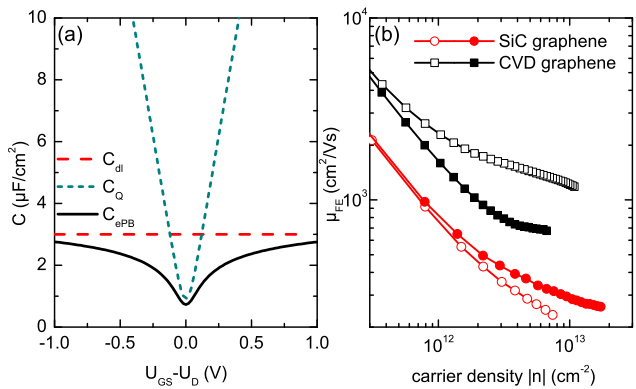}
		\caption{\label{fig:MOB} 
		(a) Capacitance of the graphene-water interface ($C_{ePB}$) calculated using an extended Poisson-Boltzmann model and its two in-series contributions: the electrolyte double-layer capacitance ($C_{dl}$) and the quantum capacitance of graphene ($C_Q$).  (b) Field effect mobility extracted from the transistor measurements using the calculated interfacial capacitance. Open and solid symbols represent p-type and n-type carriers, respectively.
		}
\end{figure}
The high transconductance of graphene SGFETs, calculated from the data in Fig. \ref{fig:TRT}(a), can be explained by the combined effects of two notable properties of graphene. The interfacial capacitance of the graphene/water system has been reported to be as high as several $\mu$F$\cdot$cm$^{-2}$.\cite{Dankerl10} In addition, the charge carrier mobilities observed in graphene are very high, even for large-scale graphene sheets.\cite{Li09} Both the interfacial capacitance and the mobility enter linearly into the transistor transconductance.
For the interfacial capacitance of the graphene/water interface, the contributions of the quantum capacitance $C_Q$ of graphene and the double layer capacitance $C_{dl}$ have to be considered. Recently, we have modeled the graphene/water interface using an extended Poisson-Boltzmann (ePB) model which considers the microscopic structure of interfacial water.\cite{Dankerl10}\\
Fig. \ref{fig:MOB}(a) shows how the total interfacial capacitance $C_{ePB}$ derived from the ePB model can be described by the series combination of $C_Q$ and a $C_{dl}$ of about 3 $\mu$F/cm$^2$. Using this model for the interfacial capacitance, the field effect mobility of charge carriers $\mu_{FE}$ in the device can be obtained after correction for the access resistance.
Fig. \ref{fig:MOB}(b) shows $\mu_{FE}$ calculated from the $I_{DS}-U_{GS}$ curves of the graphene SGFETs in Fig. \ref{fig:TRT}(a).\\
For the CVD G-SGFETs, hole mobilities greater than 1700 cm$^{2}$/V$\cdot$s are observed at the point of maximum transconductance. In the n-type region, however, the calculated mobilities are significantly lower. For the case of epitaxial graphene on SiC, electron mobilities are larger than hole mobilities. The lower carrier mobilities for the SiC G-SGFETs compared to the CVD G-SGFETs is consistent with previous publications\cite{Emtsev09,Li09}, and is attributed to the strong electronic coupling between graphene and the underlying SiC. 
Fig. \ref{fig:MOB}(b) reveals that carriers in the open channel with an opposite charge to those in the access regions show a lower mobility. That is, in the presence of the p-n junctions induced by the access regions, $\mu_{FE}$ is reduced for both CVD and SiC graphene, suggesting that the calculated $\mu_{FE}$ for carriers in the low mobility regimes (electrons in CVD graphene, holes in SiC graphene) may require correction by considering the effect of the p-n junction.\\
We further compare the graphene transistors to SGFETs based on silicon, diamond, and AlGaN/GaN heterostructures. The nitride-based devices were fabricated from GaN/AlGaN/GaN heterostructures.\cite{Steinhoff05} Diamond SGFETs were prepared using the surface conductivity of hydrogenated single crystalline diamond.\cite{Garrido05}
Fig. \ref{fig:TRT}(c) shows a comparison of the normalized transconductance $|g_m/U_{DS}|$ for all the studied SGFETs. The maximum transconductance of the CVD graphene devices is about 20 times higher than for silicon. The superior performance of G-SGFETs can be explained by the combined effect of the mobility and the interfacial capacitance, as summarized in Table \ref{tab:com}. AlGaN/GaN transistors, with mobilities similar to graphene, have a significantly lower interfacial capacitance due to the dielectric between the 2D electron gas and the electrolyte.\cite{Steinhoff05} The interfacial capacitance of the diamond devices is comparable to the graphene transistors, however, the observed hole mobilities are considerably lower. In the case of silicon SGFETs,\cite{Sprossler99} the relatively low values of capacitance and mobility result in the lowest transconductance.\\
\begin{table}
	\centering
		\begin{tabular}{|l||c|c|c|}
		  \hline
			Material&$\mu$ (cm$^2$/Vs)&$C_{int}$ ($\mu$F/cm$^2$)&$g_m/U_{DS}$ (mS/V)\\
			\hline
			Silicon&450 &0.35 &0.20 \\
			Diamond&120 &2 &0.29 \\
			AlGaN/GaN&1240 &0.32 &0.51 \\
			SiC-Graphene&400 &2 &1.14 \\
			CVD-Graphene&1700 &2 &4.23 \\\hline
		\end{tabular}
				\caption{\label{tab:com} Comparison of the materials' properties and the measured maximum transconductance for silicon,\cite{Sprossler99} diamond,\cite{Dankerl10} AlGaN/GaN,\cite{Steinhoff05} and graphene SGFETs. The values for mobility and capacitance correspond to the gate voltage where $g_m$ is maximum.
		}
\end{table}
In summary, solution-gated field effect transistors have been fabricated from large-scale graphene grown by CVD and by thermal decomposition of SiC. The transconductive sensitivity of the CVD graphene SGFETs is found to exceed 4 mS/V, almost one order of magnitude higher than for SGFETs based on other material systems, and results from the high interfacial capacitance and the large carrier mobilities in graphene. Recent progress in the growth of graphene with high carrier mobilities, together with improved device design which minimizes the access resistance, is expected to further increase the substantial advantages of graphene for sensing applications.\\
We thank A. Offenh\"{a}usser for the Si-based SGFETs and J. Howgate for the GaN-based SGFETS.  This work is funded by the German Research Foundation (DFG) in the framework of the Priority Program 1459 "Graphene", the Bavarian Graduate School CompInt, the TUM Institute for Advanced Study (TUM-IAS), and the Nanosystems Initiative Munich (NIM).

\bibliographystyle{apsrev}

\end{document}